# THE CONTRIBUTION OF TOURISM IN NATIONAL ECONOMIES: EVIDENCE OF GREECE

**Olga Kalantzi, Dimitrios Tsiotas\*, Serafeim Polyzos**
Department of Planning and Regional Development, University of Thessaly,
Pedion Areos, Volos, 38 334, Greece
*\*Corresponding author*

## ABSTRACT

*G**reece** constitutes a coastal country with a lot of geomorphologic, climatic, cultural and historic peculiarities favoring the development of many aspects of tourism. Within this framework, this article examines what are the effects of tourism in Greece and how determinative these effects are, by applying a macroscopic analysis on empirical data for the estimation of the contribution of tourism in the Greek Economy. The available data regard records of the Balance of Payments in Greece and of the major components of the Balance of the Invisible Revenues, where a measurable aspect of tourism, the Travel or Tourism Exchange, is included. At the time period of the available data (2000-2012) two events of the recent Greek history are distinguished as the most significant (the Olympic Games in the year 2004 and the economic crisis initiated in the year 2009) and their impact on the diachronic evolution in the tourism is discussed. Under an overall assessment, the analysis illustrated that tourism is a sector of the Greek economy, which is described by a significant resilience, but it seems that it has not yet been submitted to an effective developmental plan exploiting the endogenous tourism dynamics of the country, suggesting currently a promising investment of "low risk" for the economic growth of country and the exit of the economic crisis.*

## 1. Introduction

Tourism is a socioeconomic phenomenon described by the people's short-term locomotion between countries, aiming at recreation, relaxation, exploration, knowledge and generally at the emotional compensation of the travelers (Zaharatos, 1999; Dritsakis, 2004; Polyzos, 2011). The nature of tourism is by default multivariate (Po and Huang, 2008; Schubert et al., 2011; Tsiotas and Polyzos, 2014), since it refers to a set of human needs and to the corresponding economic system developed for their services, rendering nowadays this phenomenon to be considered as an emerging modern economic sector (Zaharatos, 1999). The effects of tourism are spreading in a multidimensional way, interacting, apart from the economic, also with other sectors, such as social, institutional, political, technological and environmental-ecological (Lee and Chang, 2008, Tosun et al., 2009; Polyzos et al., 2013).

Tourism suggests an important growth factor for many economies, especially for the developing ones (Lee and Chang, 2008), a fact that renders the economic approach into a significant aspect in its study, setting the relationship of tourism with economic growth as an attractive subject of research (Song et al., 2012; Panahi et al., 2014). According to the World Tourism Organization the number of international people movements globally is estimated to reach the amount of 1602 million people by 2020, while the expected tourism revenue is estimated to reach the amount of 200 billion US$ (Lee and Chang, 2008).

Under the economic perspective, tourism is an important source of income for the regions or countries of destination (the so-called tourism receptors) (Haralambopoulos and Pizam, 1996; Zaharatos, 1999; Dwyer et al., 2004; Oh, 2005; Kim et al., 2006; Lee and Chang, 2008; Po and Huang, 2008; Polyzos, 2011; Schubert et al., 2011), contributing in the improvement for such regions of their standard of living, in the increase of employment opportunities (Haralambopoulos and Pizam, 1996; Dwyer et al., 2004; Galani - Moutafi, 2004; Oh, 2005; Kim et al., 2006; Lee and Chang, 2008; Po and Huang, 2008; Polyzos, 2011; Schubert et al., 2011), in the provision of capital for new investments and in the development of infrastructures (Buhalis, 2001; Dritsakis, 2004; Polyzos, 2011; Schubert et al., 2011; Song et al., 2012), in regional growth (Lee and Chang, 2008; Polyzos, 2011; Song et al. 2012; Polyzos et al., 2013; Panahi et al., 2014), in the import of tourist exchange (Oh, 2005; Kim et al., 2006; Po and Huang, 2008; Schubert et al., 2011; Panahi et al., 2014) and in the balance of payments (Meade, 1951; Oh, 2005; Kim et al., 2006; Polyzos, 2011).

In the case of Greece, the scientific documentation of tourism commenced after the Second World War, while major tourism development started in the middle of '70s, taking advantage of the that times' unpopularity of Spanish resorts, which stimulated the demand for alternative Mediterranean destinations. Tourism flows immensely increased in Greece in the '80s, facilitated by plenty of natural, cultural and environmental resources, along with the accessibility to major islands and with the relatively lower cost of living, comparatively to the European competition. This background rendered Greek tourism capable in satisfying a great diversity of tourism demand and consequently the tourism industry to grow rapidly emerging into a significant developmental factor of the national economy (Buhalis, 2001).

Greek tourism is included among the biggest economic activities, which contributes in an amount of 17% to the *Gross National Product* (GNP) and it creates a total demand of 34 billion Euros (€) in the national economy. For the year 2013 the tourism income reached almost the amount of 12 billion €, covering the 51% of deficit of the *Trade Balance* (TB). Greek tourism occupies in total (directly and indirectly) about 800.000 workers, a number that corresponds in the 18,3% of the national employment, the bigger part of which is activated in regional scale, in small- or medium-level enterprises (AGTE, 2014).

Generally, the performance of Greek tourism is particularly satisfactory, relatively to the country's size, highlighting the importance of tourism for the national and regional economic growth. According to the World Tourism Organization (UNWTO), Greece in 2012 occupied the 17$^{th}$ place in the ordering of international arrivals and the 23$^{rd}$ in the level of income. Also, Greece in 2013 occupied the 32$^{nd}$ place, among 140 countries, in the scores of the Travel and Tourism Competitiveness Indicator, while its General Competitiveness Indicator is ranked hardly in the 96$^{th}$ place. This implies that the Greek tourism suggests a fundamental sector of the national economy, which is worldwide competitive (AGTE, 2012, 2014; UNWTO, 2014; WTTC, 2014).

The major characteristic of tourism in Greece, along with other - mainly Mediterranean - countries, regards its "intensity" (mass tourism), due to the income increase and to the improvement of the prosperity level in the countries of origin. Before the appearance of the mass tourism, vacations of cultural or recreational character were an exclusive privilege of a minority of rich sightseers. The massive character of tourism in Greece and in most Mediterranean countries is based on the promotion of the "3*S*" (*sea*, *sun*, *sand*), in combination with the cultural heritage, and it is mainly developed in the coastal and island regions at the summer period (Galani-Moutafi, 2004; Polyzos, 2011; Eeckels et al., 2012).

Another characteristic of the Greek tourism concerns its spatial and seasonal intensity, inducing significant negative effects in the quality of the natural and urban environment, as well as in the regional societies and economies (Galani - Moutafi, 2004; Salvati et al., 2014). Apart from the tourism's positive contribution in the economy and employment, the environmental degradation due to mass tourism has already started to appear in Greece from the '70s, electing the risks that may appear due to an touristic area's excessive economic dependence on the tourism activities (Polyzos and Tsiotas, 2012; Salvati et al., 2014b).

Within this framework, this article studies the effects that tourism induced in these sectors in Greece, in the period 2000-2012. In particular, Section 2 presents the theoretical-conceptual framework of the economic effects of tourism and its determinative factors, Section 3 analyzes the effects in the national income, employment, regional development and tourist exchange.  Moreover, Section 3 evaluates the importance of tourist exchange in comparison to the other invisible revenues (or invisible resources) estimating the contribution of the former in the deficits of the trade balance of the country. Finally, in Section 4 conclusions are formulated based on the previous part of the analysis.

**2. Aspects of the tourism contribution in the national economy**

Tourism is considered today as an increasing overall economic activity with positive effects on the increase of long-run economic growth and, according to this perspective, it is encountered as a desirable phenomenon that contributes positively to the development of the national economy of the receptor countries (Dwyer et al., 2004; Schubert et al., 2011). The economic changes induced from tourism to the receptor countries can be classified into the following basic categories (Polyzos, 2011):

*2.1. Increase of Income*
According to the supply and demand theory, the major factors determining the budget line of a tourist are the price of the tourism good and the income of the tourist. In most of the empirical tourism demand studies, the income of the country or region of origin, the cost of visiting a destination and the substitute prices of alternative destinations are the most commonly considered determinants. Through this perspective, tourism contributes in the increase of income of those who directly or indirectly are occupied in the sector, since it constitutes a demand's source of goods and services (such as transport, accommodation, food, clothing, entertainment, cultural goods, etc). This increase is related to the level of tourist demand and the total consuming expense per tourist (Haralambopoulos and Pizam, 1996; Zaharatos, 1999; Dwyer et al., 2004; Oh, 2005; Kim et al., 2006; Lee and Chang, 2008; Po and Huang, 2008; Schubert et al., 2011).

The tourist consumption strengthens the national income of the state and it helps to the solution of the budgetary problem. The contribution of tourism is of great importance in the increase of the regional (municipal) income, since the local authorities of the tourist receptor regions increase their income, through the imposition of taxes for the usage of municipal or other infrastructures and for the exploitation of the public property. Evaluating the economic effects of tourism based on their "direct impact", we can classify them into *direct*, *indirect* and *induced* effects (Schubert et al., 2011), according to the framework illustrated in diagram 1.

Direct effects are a result of tourism expenditure and they influence the enterprises that the biggest part of their economic activities is related to the tourism market. Such enterprises are hotels, restaurants and food services, trade, transport, entertainment and culture firms. Changes in the level of production due to tourist consumption, in conjunction with the increase of the raw inputs of such enterprises, they represent the direct effect of tourism.

The enterprises of the first class, in order to satisfy the tourist demand, they get a considerable part of their supplies from other enterprises. Consequently, the tourist demand indirectly influences some other sectors, such as foods, drinks, marketing, maintenance of transportation means, legal services, accountant services, etc. This procedure sequentially induces indirect effects, which concern the enterprises that are connected indirectly with the tourist sector and constitute the suppliers of consuming goods of the enterprises that are directly involved with tourism.

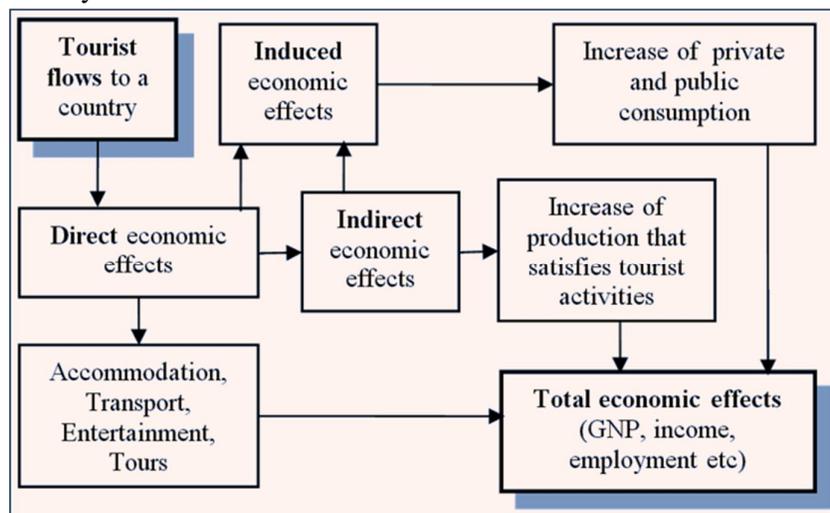

**Diagram 1:** Classification of the economic effects of tourism, based on their direct impact (own elaboration).

Both the direct and indirect effects provide incomes to those who work for the satisfaction of tourist demand. These incomes also affect the regions constituting the tourism receptors, influencing the total consumption of households and the final demand for goods and services. Such impacts are called *induced effects* of tourism. The sum of the *direct* ($e_1$), *indirect* ($e_2$) and *induced* effects ($e_3$) gives the amount of the *total effects*, according to relation (1), shaping the total changes of the basic economic measures, such as the GNP, the income, the employment, etc.

$$Total\ Effects = E_{tot} = \sum_{i=1}^{3} e_i \qquad (1)$$

Finally, one aspect of the negative criticism that mass tourism is submitted to is due to the fact that the majority of the tourist income does not remain in the tourist destination, but it "leaks" to other regions. This is a result of massiveness that mainly favors the operation of enterprises of large-scale, which they do not use locally produced products, but goods imported from other regions (Polyzos, 2011).

*2.2. Increase of Employment*

Tourism induces employment opportunities to the regional (local) labor and, more generally, to those who involve with the productivity chain of tourism. Employment opportunities emerge both for these that are directly connected with the tourism sector (direct employment) and as also for those that are indirectly connected (indirect employment), by providing to tourists either goods or services (Haralambopoulos and Pizam, 1996; Dwyer et al., 2004; Galani - Moutafi, 2004; Oh, 2005; Kim et al., 2006; Lee and Chang, 2008; Po and Huang, 2008; Polyzos, 2011; Schubert et al., 2011).

Moreover, tourism labor often does not originate from the region of residence, but from other regions and many times from other countries. Thus, the form of tourism favors an important sum of tourism demand to be satisfied by other regions, resulting to the "leak" of a considerable amount of money that otherwise it would help in the improvement of the standard of living of the local societies. On the contrary, the alternative modes of tourism provide opportunities to the local population for occupying in the tourist market and they also favor the local production of goods, resulting more to the endogenous satisfaction of the consumption demand and to the increase of the tourism's contribution in the regional development (Polyzos, 2011)

*2.3. Regional Development*

The increased economic activity induced by tourism provides opportunities for regional development of the receptor countries. The regional or national multipliers capture the changes in the income, the employment and the other measures of the regional or national economy, induced by a unitary increase of the demand. Trade transactions lead a part of the demand out of the region or country, resulting to the reduction of the relative multiplier. Both the structure of final demand and the scores of multipliers of a tourist region influence the final result induced by the increase of income, employment, etc. As much as bigger is the rate of satisfaction of the final demand from the local market, so much bigger is the size of the multiplier and thus more positive are the changes in the economy of the tourist region (Lee and Chang, 2008; Polyzos, 2011; Song et al. 2012; Polyzos et al., 2013).

A major concern from this perspective is the seasonal concentration of tourist movement in few months per year (diagram 2), which causes serious negative effects in all the aspects of the tourist circle. This affects the physical, economic, cultural, structured and anthropogenic environment, inducing greater pressures in the peak months, reaching many times the boundaries of the carrying capacity of the tourism receptors (Polyzos, 2011; Polyzos et al., 2013).

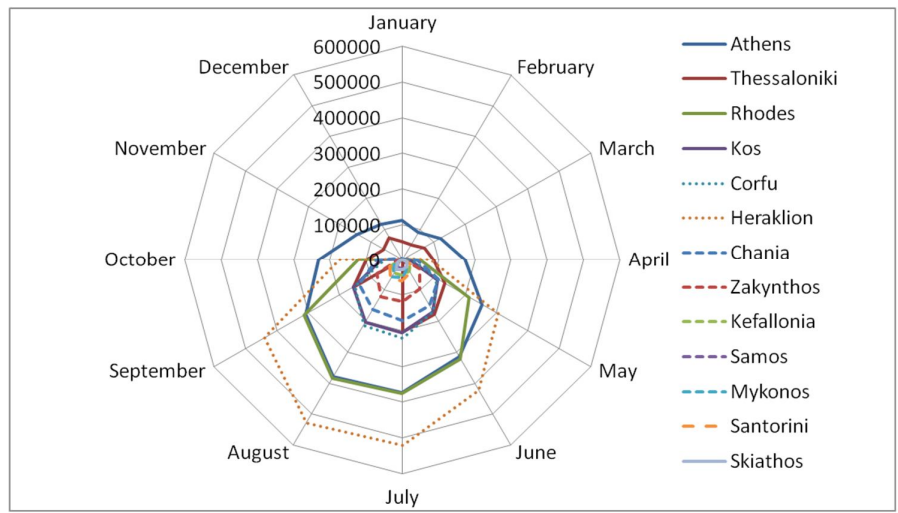

**Diagram 2:** Seasonal change of tourist arrivals in the basic Greek airports for the year 2013 (source: BOG, 2014; own elaboration).

Seasonality causes functionality problems to tourist enterprises, where, in the majority of the Greek insular regions, over the 50% of the total annual touristic activity is conducted within the 3-month period July–September, inducing the inactivation of the invested tourism capital for over 6 months.

Consequently, the tourist enterprises are overloaded with high standard expenses, in order to face their functional needs during the peak period, a fact that increases the average operational cost and it decreases the total profitability. Finally, seasonality also negatively influences the medium size firms, since they are compelled to possess equipment that is operational annually only for a short time period and it stays inactive for the other time without participating to the productivity procedure (Polyzos et al., 2013; Tsiotas and Polyzos, 2014).

*2.4. Import of Tourist Exchange*
Foreign tourism reinforces the import of exchange in the receptor country or region, which in turn leads to economic growth (Oh, 2005; Kim et al., 2006; Po and Huang, 2008; Schubert et al., 2011; Panahi et al., 2014). This contributes in the improvement of *Balance Of Payments* (BP) and in the regional growth. The BP is of particular importance for the economy of a country, since it influences the configuration of economic measures, such as the national income and the national expenses, while simultaneously it portrays the economic dynamics of the receptor country (Meade, 1951; Oh, 2005; Kim et al., 2006).

The amount of tourist expenditure refers to the amount of money that each tourist spends during the staying in the tourist destination and it is directly connected with the tourism's contribution in the economic development. The tourism expenditure significantly depends on the welfare of the country of origin, implying that tourists originating from countries with high standards of living and high per capita incomes usually spend greater amounts in their tourist vacations. Additionally, the income class of tourist is of great importance for the amount of tourist expenditure, since wealthy tourists spend greater amounts in comparison to others of lower incomes (Oh, 2005; Kim et al., 2006; Po and Huang, 2008; Schubert et al., 2011).

Finally, apart from the amount tourists spend during their vacations in the tourist destination, the composition of final demand is of particular importance for the local and regional growth. In the case that the tourist demand is being satisfied by local products and goods, then tourism contributes more in the regional and local development than in the opposite case, where imported goods satisfy the tourist demand (Oh, 2005; Kim et al., 2006; Po and Huang, 2008; Schubert et al., 2011).

**3. Methodological Framework**
This paper applies a methodology for the macroscopic study of the tourism contribution to the Greek economy, consisting of a descriptive and an empirical part of analysis, as it is shown in diagram 3. The first part classifies the impacts of tourism according to the previous literature consideration and examines their applicability on the Greek economy, based on diachronic statistical data from the period 2000-2012.

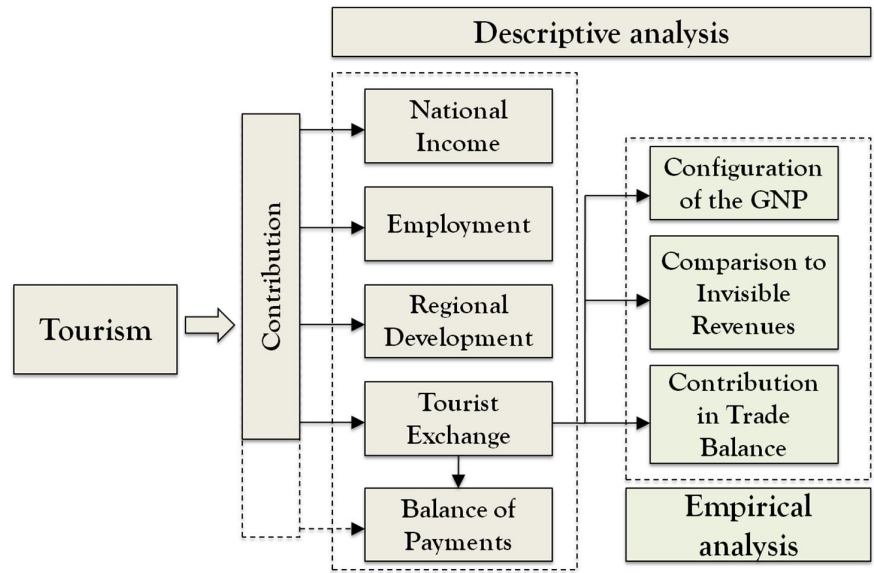

**Diagram 3:** The methodological framework of the analysis (source: BOG, 2014; own elaboration).

The empirical part of the analysis focuses on the diachronic variable of *tourism* or *travel exchange* (TE) and examines its contribution to the *balance of payments* account (BP), using statistical techniques. This perspective is adopted because tourist exchange operates as an alternative form of exports (Oh, 2005; Kim et al., 2006) and thus it can contribute to the BP through foreign exchange earnings and proceeds that they are generated from tourism expansion, suggesting a significant source of income for the national economy of Greece.

In particular, the quantitative methods used in the analysis are the independent samples *t*-test for the comparison of means, the *Pearson*'s *bivariate coefficients of correlation* and the *simple linear regression* model. First, the independent-samples *t*-test (Hays, 1981; Norusis, 2004) is used for the comparison of the means $\mu_1$ and $\mu_2$ between two groups $\{G_1, G_2\}$ of cases that originate from the same initial set. The procedure utilizes the Levene's test for the equality of variances and produces a separate hypothesis *t*-testing result per case (assuming separate/unpooled and pooled variances), where the proper one is chosen according to its significance. In this test, the *t*-values and the degrees of freedom (*df*) are calculated according to the type of variance (unpooled or pooled), while the significances are obtained from the *t* distribution separately for each of the *t*-values (Norusis, 2004).

Next, the Pearson's bivariate coefficient of correlation (Walpole et al., 2012; Tsiotas and Polyzos, 2014, 2015) is used for capturing linear relations between pairs of variables. Its formula is shown at relation (1), where $cov(x,y) \equiv s_{xy}$ stands for the covariance of the variables $x,y$ and $s_x$, $s_y$ for their sample standard deviations, respectively. The coefficient $r_{xy}$ ranges within the interval [-1,1], indicating a perfect positive linear relation when $r_{xy}$=1 and negative when $r_{xy}$=-1.

$$r(x,y) \equiv r_{xy} = \frac{\operatorname{cov}(x,y)}{\sqrt{\operatorname{var}(x) \cdot \operatorname{var}(y)}} = \frac{s_{xy}}{s_x \cdot s_y} \qquad (2)$$

The rationale for applying the correlation analysis is to detect structural similarities in the diachronic fluctuations (2000-2012) of the examined variables. Variables with strong linear relations ($|r_{xy}| \sim 1$) they fluctuate similarly in the time line and thus they are supposed to have better structural relevance than others.

Finally, the curve fitting procedure is used for quantifying a linear patter $y=bx+c$ that describes the diachronic fluctuation of the examined variables. The existence of a statistically significant coefficient beta (*b*) indicates an evolving trend (either growing or decaying) for the response variable *y*, while the coefficient of determination $R^2$ (*R*-square) describes that an amount of $100 \cdot R^2$ % of the variation of the response variable's (*y*) data are captured by the linear model (Norusis, 2004, Tsiotas and Polyzos, 2014, 2015).

**4. The contribution of Tourism in the Greek Economy**
*4.1. Contribution to the National Income*
One of the most important effects of tourism is its contribution in the increase of national income, suggesting a multi-factorial procedure. Tourism contributes in the incomes' increase of the occupations that are directly or indirectly related to the tourist sector, since it induces an increase of demand of goods and services in the -relevant to tourism- economic activities, such as transportation (Tsiotas and Polyzos, 2014), accommodation, diet, clothing, entertainment, culture, and also in the interrelated to tourism economic sectors (Haralambopoulos and Pizam, 1996; Zaharatos, 1999; Dwyer et al., 2004; Oh, 2005; Kim et al., 2006; Lee and Chang, 2008; Po and Huang, 2008; Polyzos, 2011; Schubert et al., 2011). According to Dritsakis (2004), a point of interest for the regional development is the wide distribution of the tourist income in the population of country and the increase of income for the residents of the less developed regions.

Tourism income in Greece is 2.5 times greater than the industrial income and 1.8 times greater than the total profit of exports. However, these figures miss to capture the pre-purchases of currency by tourists abroad, the credit card payments or the payments for cruises and other earnings, which they obviously increase the total receipts of tourism up to 80% more (Buhalis, 2001). According to the available statistics (BOG, 2014), the tourism income possesses a major proportion in the total of the exchange income of Greece. For the period 2000-2013 the tourist exchange of the country reached the amount of 150 billion €, when the total income from exports for the corresponding period reached the amount of 221 billion €. This illustrates that the income from the tourist exchange is of similar dynamics with that of imports, implying that the former is capable to cover almost the total the deficit that imports induce in the country.

Another indication of the significant effect that the tourist activity induces to the national income derives from the contribution of tourism in the GNP of country, which is recorded 16.5%, 16.4% and 16.3% for the years 2011, 2012 and 2013 respectively (AGTE, 2014). Additionally, the Foundation for Economic and Industrial Research (FEIR, 2012) estimated that the total effect of tourism for the year 2012 reached the amount of 34.4 billion €, participating at an amount of 15,1% in the total GNP of the country, implying that tourism constitutes the country's "heavy industry" and that it recommends a basic pylon of the national economy.

*4.2. Contribution to Employment*

Tourism creates employment opportunities for the receptor countries, favoring the occupations that are both directly (direct employment) or indirectly (indirect employment, which provides in the tourist sector consuming products and services) connected with the tourist sector (Lee and Chang, 2008; Polyzos, 2011). In Greece, the total employment (direct and indirect) in the tourist sector reached indicatively, for the year 2006, the amount of 867 thousand places of work, corresponding approximately to the 20% of the total employment (Eeckels et al., 2012). Also, it is estimated that employment in the tourism sector in Greece reaches the amount of 10% (6.1% corresponds to direct employment and 3.9% to indirect) of the total employment in the country (Buhalis, 2001)

In particular, the participation of tourism in the direct and total employment in Greece, for period 2000-2012, is shown in table 1.

**Table 1**

Travel & Tourism Contribution to Employment

| Year | **Direct Contribution**[(*)] | **Total** (Direct+Indirect) **Contribution**[(*)] |
|---|---|---|
| 2000 | 8.5 | 19.2 |
| 2001 | 8.3 | 19.4 |
| 2002 | 8.1 | 19 |
| 2003 | 7.6 | 18.1 |
| 2004 | 7.5 | 18.2 |
| 2005 | 8.2 | 19.5 |
| 2006 | 8.3 | 19.8 |
| 2007 | 8.1 | 19.4 |
| 2008 | 7.9 | 18.7 |
| 2009 | 7.3 | 17.7 |
| 2010 | 7.8 | 17.8 |
| 2011 | 8 | 17.4 |
| 2012 | 8.1 | 17 |

(*) % participation
(source: WTTC, 2014; own elaboration)

According to table 1, tourism in Greece considerably contributes in the creation of employment, in an amount that ranges to 17-19.8% (total employment), where an amount of 7.3-8.5% refers to the direct and of 9.7-11.3% to the indirect employment. It is remarkable that the recorded decrease of the total employment in tourism, which is initiated from the year 2009, it does not exceed two percentage units (2%), implying the resilience of tourism, whether interpreting the latter as an economic aspect and taking under consideration the economic crisis (Polyzos et al., 2013; Tsiotas and Polyzos, 2014) that Greece is currently submitted to.

According to the annual report of World Travel and Tourism Council (WTTC, 2014), the participation of tourism in the Greek employment is estimated that it presents increasing trends and that it is about to reach the year 2024 the amount of 21.4% of the total employment, supporting about 936 thousand places of work. This estimation seems to exceed the local maximum of 19.8%, presented in the year 2006 (table 1), a fact that verifies the aforementioned observation about the resilience and the dynamics of tourism in Greece.

*4.3. Contribution to Regional Development*

Tourism is closely related to regional growth, since the exploitation of the tourist resources of each region favors the increase of tourist flows and the consecutive economic regional growth. Empirical researches showed that tourism growth is favored in receptor societies that are submitted to a transitive stage of development, where the other components of their economy are subjected to depression (Lee and Chang, 2008; Polyzos, 2011; Song et al. 2012; Polyzos et al., 2013). Particularly for the case of Greece, the intensity of tourism traffic seems to be related to the geographic distribution of its cultural resources, but also with its natural resources and amenities. This structure emerges the spatial dimension of tourism (Tsiotas and Polyzos, 2014) and illustrates the relation of the latter with the regional development (Polyzos, 2011).

Besides, the construction of infrastructures for tourist activities, such as units of hospitality, of feeding and other supplementary activities (athletic, cultural, etc), the creation of transportation stations and the upgrade of transportation networks (Tsiotas and Polyzos, 2014, 2015), which aim in the service of the tourism activities, all the above considerably contribute in the regional economic development (FTR, 2014). An indicative example to this suggests the development of integrated tourist destination units (ITDA) in the region of Messenia, in the western Peloponnese (Kalantzi and Tsiotas, 2010), which contributed in the tourist development of this region.

In the work of Polyzos et al. (2013) the authors noted that the Greek regions that reach the limits of saturation in their tourist growth have a geographic relevance, an observation that accredits a previous statement about the tourism dynamics of the receptor societies that are submitted to a transitive stage of development. In particular, the authors observed that the Aegean regions are described by higher coefficients of saturation in comparison to the corresponding Ionian regions, but they also made the impressive and unexpected observation that the metropolitan regions of Attiki and Thessaloniki have not yet reached their limits of saturation. This consideration implies that tourism in Greece maps an unbalanced geographical distribution, a fact that necessitates the application of an effective regional policy, in order to regulate the development of the Greek regions under the criterion of equality.

*4.4. Contribution to Tourist Exchange*
The term "travel of tourist exchange" defines the initial monetary expense that enters, through the tourist consumption, in a receptor country (Oh, 2005; Kim et al., 2006; Po and Huang, 2008; Schubert et al., 2011). Tourism constitutes a phenomenon that, apart from the social acts of locomotion and social interaction, it has a significant economic impact, contributing in the flow of (money) exchange from the country of origin to the country of destination. This suggests a vital process for the national economy of the receptor country, which is of particular importance in the cases of the developing countries (Lee and Chang, 2008; Tosun et al., 2009; Song et al., 2012), as far as it concerns the flow of real and not nominal (accountant) amounts of money in the domestic market, which are directly exploitable.

The previous consideration suggests an issue of great importance for the case of Greece, and particularly at the current period of the unconventional model of the Greek economic crisis (Tsiotas and Polyzos, 2014), where the country is seeking exits based on its healthy productive mechanisms, where tourism suggests one of them. Table 2 presents the diachronic evolution of the tourist exchange in Greece, at the period 2000-2012, illustrating a distinctive decrease at the year of entering the crisis 2009 and the next year 2010.

**Table 2**
Diachronic evolution of tourist exchange in Greece

| Year | Travel – Tourism Exchange (billion €) |
|---|---|
| 2000 | 10.061 |
| 2001 | 10.58 |
| 2002 | 10.285 |
| 2003 | 9.495 |
| 2004 | 10.348 |
| 2005 | 10.73 |
| 2006 | 11.357 |
| 2007 | 11.319 |
| 2008 | 11.636 |
| 2009 | 10.4 |
| 2010 | 9.611 |
| 2011 | 10.505 |
| 2012 | 10.443 |

(Sources: WTTC, 2014; BOG, 2014; own elaboration)

According to table 2, the scores of tourist exchange in the years 2009 and 2010 reflect the impact of the national economic crisis in the tourist sector, a fact which presents a tendency of recovering at the next couple of years 2011-2012. This becomes clearer in the diagram 4 that shows the diachronic evolution of tourist exchange, based on the available data shown in table 2. On the one hand, this observation obviously outlines the negative effect of the economic crisis on tourism, illustrating a seven-year backspace that was enforced to the tourist growth (Greece in 2010 returned to the levels of tourist exchange recorded in the year 2003).

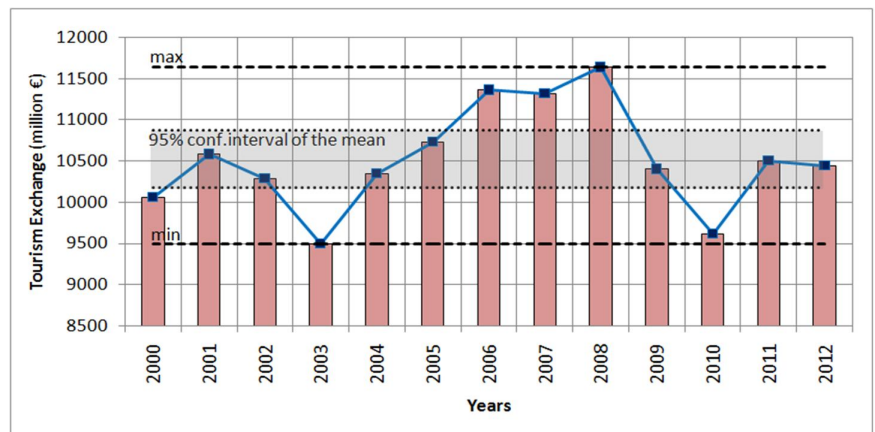

**Diagram 4:** Diachronic evolution of tourist exchange (source: BOG, 2014; own elaboration).

On the other hand, the performance of Greece in the next years 2011-2012 illustrate that tourism is described by a resilient homoeostatic mechanism, which quickly activated the recovering procedure that restrained the duration of the forcible bending due to the economic crisis for hardly two years.

As a further consideration, the diagram 5 is constructed on data available from the Foundation for Tourism Research - FTR (2014), depicting the relative amount of the tourism expenditures for 12 countries of origin, which they have chosen Greece as their tourism destination, for the year 2013.

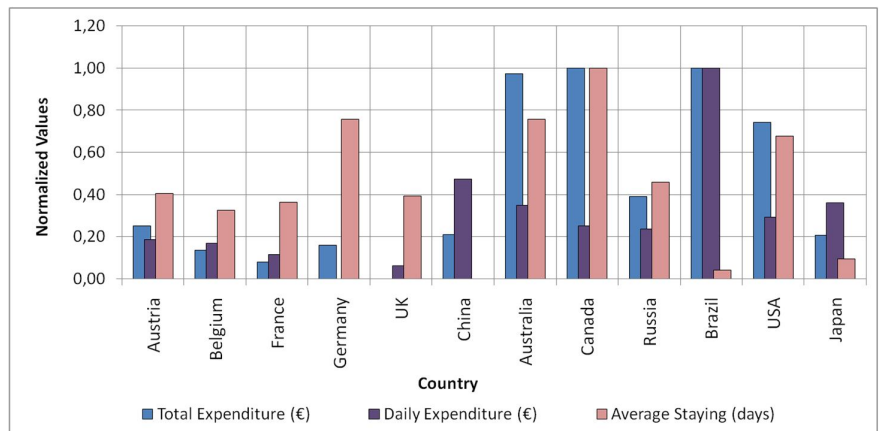

**Diagram 5:** Normalized scores (relative amount) of the tourism expenditure spent in Greece per country of origin, for the year 2013 (source: FTR, 2014; own elaboration).

According to diagram 5, tourists from Canada and Brazil spent the greatest total amount for their vacations in Greece. Nevertheless, Brazilian tourists have almost three times higher daily expenditures than the Canadian, implying that they prefer to stay less days for vacations enjoying services of higher quality than the latter. The decomposition of the total expenditure into the daily expenditure and into the average of staying, as it is shown in diagram 5, it may provide useful insights about the tourism preferences, or the "tourism cultivation", or perhaps about the income class of the foreign tourists visiting a receptor country. In the case of Greece, tourists originating from Germany and next from the UK have longer vacations, but they spend the smallest amounts of daily expenditure, implying that they prefer to enjoy basic tourism services but for a longer time period or that they are probably originating from middle or lower income layers comparatively with the standards of their country.

*4.5. Reduction of the deficit in the Balance of Payments*
One of the most important effects of international tourism in the national economy of a receptor country is the contribution of the former in the *Balance of Payments (*BP) (Oh, 2005), a fact that becomes of even greater importance in the case of developing countries (Zaharatos, 1999; Tosun et al., 2009; Polyzos, 2011; Song et al., 2012). The BP is defined as the international special account that records the size and the evolution of economic transactions conducted by a country worldwide in comparison to the others. This account includes records of the bipartite capital interactions (inputs – outputs) in monetary exchange between countries, for a given time period (Zaharatos, 1999; Polyzos, 2011).

The preserved deficit of the *Trade Balance* (TB) in Greece is diachronically covered by the balance of the *Invisible Revenues* (IR), in which tourism suggests an important economic component (Polyzos, 2011; BOG, 2014). This is examined in more detail in the next sub-sections on available statistical data referring to the BP of Greece.

4.5.1. The contribution of Tourist Exchange in the configuration of the GNP
The diachronic evolution of tourist exchange flowing into a receptor country suggests an important indicator for the tourist dynamics of this country (Zaharatos, 1999). Table 3 provides relevant insights with this statement, presenting the diachronic change of the tourist exchange in Greece, in comparison to the GNP, for the years 2000-2012. According to this table, the contribution of the tourist exchange in the GNP of the country, at the period 2000-2012, is ranged in a percentage between 4.33-7.38%.

For the illustration of the information provided in table 3, the diagrams 6 and 7 are examined.

**Table 3**

Diachronic change of tourist exchange in Greece

| Year | GNP | Travel – Tourism Exchange | Participation to the GNP |
|---|---|---|---|
| | | (billion €) | (%) |
| 2000 | 136.3 | 10.061 | 7.38 |
| 2001 | 146.4 | 10.58 | 7.23 |
| 2002 | 156.6 | 10.285 | 6.57 |
| 2003 | 172.4 | 9.495 | 5.51 |
| 2004 | 185.3 | 10.348 | 5.58 |
| 2005 | 193 | 10.73 | 5.56 |
| 2006 | 208.6 | 11.357 | 5.44 |
| 2007 | 223.2 | 11.319 | 5.07 |
| 2008 | 233.2 | 11.636 | 4.99 |
| 2009 | 231.1 | 10.4 | 4.50 |
| 2010 | 222.2 | 9.611 | 4.33 |
| 2011 | 208.5 | 10.505 | 5.04 |
| 2012 | 193.3 | 10.443 | 5.40 |

(Sources: WTTC, 2014; BOG, 2014; own elaboration)

First, diagram 6 shows the relation of the tourist exchange (TE) in accordance to the GNP in Greece, constructed on values of the 13 year period 2000-2012. According to this diagram, the relation TE = *f*(GNP) can be considered with a 50.6% precision as linear, indicating that these variables (TE, GNP) are positively correlated. This implies that a increase of the tourist exchange induces a corresponding increase in the GNP of the country.

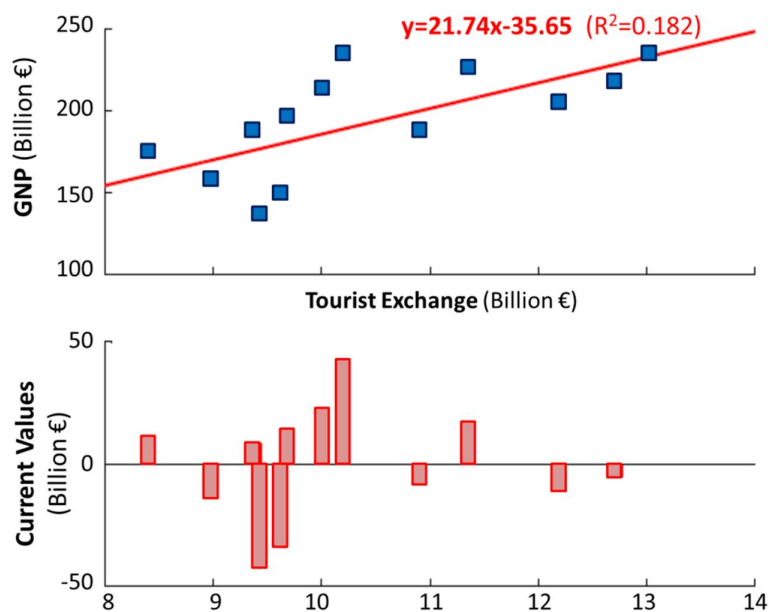

**Diagram 6:** Change of tourist exchange in comparison to GNP (data from table 2; own elaboration)

Further, diagram 7 constitutes a comparative graph that presents in absolute values how the relation between TE and GNP diachronically evolves. According to this diagram, the diachronic evolution of the tourist exchange sketches a more invariant time curve in comparison to the respective of GNP, a fact that accredits the consideration that tourism suggests a fundamental component of the national economy and moreover a factor of economic growth in Greece.

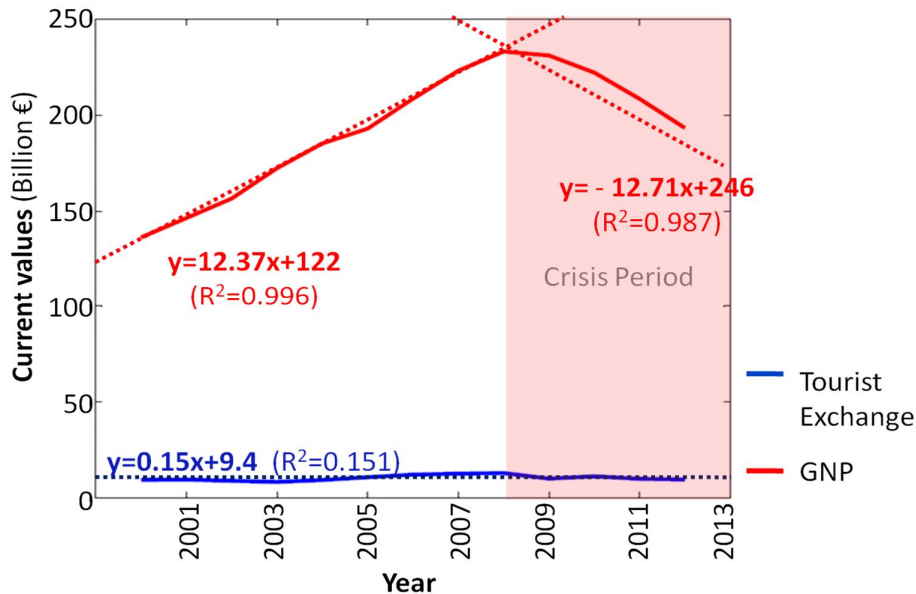

**Diagram 7:** Annual change of tourist exchange, in accordance to GNP. The colored area corresponds to the crisis period (data from table 3; own elaboration)

Diagram 7 also illustrates that the rapid increase in the GNP of Greece, which is observed at the period 2000-2008, mainly suggests an effect of productivity components that are unrelated to the tourism, which they were proved to be unstable developmental factors for the period of the economic crisis.

For example, it seems that the conduct of the Olympic Games in Athens, in the year 2004, it did not contribute in the expected increase of tourism in Greece, since the tourist arrivals of both the foreign and domestic tourism were reduced (BOG, 2005). Whatever was the relevant increase in the domestic demand the years prior to 2004, it was probably related more to the investment programs referring to the preparation of the Olympic Games than to the tourism motivation itself (BOG, 2005).

4.5.2. The Invisible Revenues and the Balance of Invisible Resources

In the *Balance of the Invisible Revenues* (BIR) are recorded the income and payments from sources that are not as obvious as those of Trade Balance (exports - imports of goods). The term of *Invisible Resources* (IR) defines the income and expenses from services, such as tourism and transports, the income and the payments of interests, dividends, wages and pensions, as well as the incoming and outgoing transfers, such as those from the EU and the migratory transfers. According to the recent method of recording the transactions in the Balance of Payments, which is applied from 1999 for complying to the recommendation of the *International Monetary Fund* (IMF), the BIR consists of three separate accounts, the *Balance of Services* (BS), the *Balance of Incomes* (BI) and the *Balance of Current* (BCT) and *Capital Transfers* (BCAT) (ECB, 2007).

The diachronic evolution of the BIR in Greece, for the period 2000-2012, is shown in table 4. According to this table, the BIR is firmly positive and with important increasing tendencies up to the year 2008. In the year 2009, where Greece entered the economic crisis, an abrupt decrease of the BIR is observed, which is however stopped the next years, remaining since in the levels of 2009.

**Table 4**

Diachronic change of the invisible revenues balance

| Year | $C$ | $C_1$ | $C_1/C$ | $C_2$ | $C_2/C$ | $C_3$ | $C_3/C$ | $C_4$ | $C_4/C$ |
|---|---|---|---|---|---|---|---|---|---|
| 2000 | 32.14 | 20.98 | 65.28 | 3.04 | 9.46 | 5.74 | 17.86 | 2.39 | 7.44 |
| 2001 | 33.12 | 22.08 | 66.67 | 2.10 | 6.34 | 6.34 | 19.14 | 2.60 | 7.85 |
| 2002 | 30.56 | 21.13 | 69.14 | 1.63 | 5.33 | 5.98 | 19.57 | 1.82 | 5.96 |
| 2003 | 31.83 | 21.43 | 67.33 | 2.58 | 8.11 | 6.42 | 20.17 | 1.39 | 4.37 |
| 2004 | 38.53 | 26.74 | 69.40 | 2.81 | 7.29 | 6.36 | 16.51 | 2.62 | 6.80 |
| 2005 | 39.73 | 27.25 | 68.59 | 3.27 | 8.23 | 6.88 | 17.32 | 2.32 | 5.84 |
| 2006 | 42.05 | 28.36 | 67.44 | 3.53 | 8.39 | 6.85 | 16.29 | 3.31 | 7.87 |
| 2007 | 47.18 | 31.34 | 66.43 | 4.56 | 9.67 | 6.61 | 14.01 | 4.67 | 9.90 |
| 2008 | 51.16 | 34.07 | 66.59 | 5.57 | 10.89 | 6.88 | 13.45 | 4.64 | 9.07 |
| 2009 | 38.98 | 26.98 | 69.21 | 4.28 | 10.98 | 5.38 | 13.80 | 2.33 | 5.98 |
| 2010 | 39.50 | 28.48 | 72.10 | 4.01 | 10.15 | 4.65 | 11.77 | 2.36 | 5.97 |
| 2011 | 39.30 | 28.61 | 72.80 | 3.32 | 8.45 | 4.44 | 11.30 | 2.93 | 7.46 |
| 2012 | 39.05 | 27.53 | 70.50 | 3.83 | 9.81 | 5.13 | 13.14 | 2.56 | 6.56 |

$C$. Total of Invisible Revenues – IR (billion €)  
$C_1$. Balance of Services – BS (billion €)  
$C_1/C$ . Contribution of the BS in the balance of IR (%)  
$C_2$. Balance of Income – BI (billion €)  
$C_2/C$ . Contribution of the BI in the balance of IR (%)  
$C_3$. Balance of Current Transfers - BCT (billion €)  
$C_3/C$ . Contribution of the BCT in the balance of IR (%)  
$C_4$. Balance of Capital Transfers - BCAT (billion €)  
$C_4/C$ . Contribution of the BCAT in the balance of IR (%)  
(source: BOG, 2014; own elaboration)

However, it is important to note that at this period the BS was maintained in satisfactory levels and that it was submitted to a relatively insignificant decrease, a fact that illustrates a diachronically invariant performance for this account, where one of its basic components is the Tourist Exchange.

4.5.3. The importance of Tourist Exchange in comparison to the other Invisible Revenues

As it was observed in table 4, the BS occupies the lion's share in the BIR, where its contribution ranges between 65-73%. Taking under consideration that the Travel or Tourist Exchange (TE) suggests a fundamental component of the BS, the participation of the former to the configuration of the latter is expected to be significant. For evincing this consideration, the rates of travelling exchange, in comparison to the corresponding of the BIR and to each of the IR's components, are shown in table 5.

**Table 5**

Diachronic change of tourist exchange in comparison to the change of the invisible revenues

| Year | $C_{11}$ | $C$ | $C_{11}/C$ | $C_1$ | $C_{11}/C_1$ | $C_2$ | $C_{11}/C_2$ | $C_3$ | $C_{11}/C_3$ | $C_4$ | $C_{11}/C_4$ |
|---|---|---|---|---|---|---|---|---|---|---|---|
| 2000 | 10.1 | 32.1 | 31.46 | 21.0 | 48.10 | 3 | 336.67 | 5.7 | 177.19 | 2.4 | 420.83 |
| 2001 | 10.6 | 33.1 | 32.02 | 22.1 | 47.96 | 2.1 | 504.76 | 6.3 | 168.25 | 2.6 | 407.69 |
| 2002 | 10.3 | 30.6 | 33.66 | 21.1 | 48.82 | 1.6 | 643.75 | 6 | 171.67 | 1.8 | 572.22 |
| 2003 | 9.5 | 31.8 | 29.87 | 21.4 | 44.39 | 2.6 | 365.38 | 6.4 | 148.44 | 1.4 | 678.57 |
| 2004 | 10.3 | 38.5 | 26.75 | 26.7 | 38.58 | 2.8 | 367.86 | 6.4 | 160.94 | 2.6 | 396.15 |
| 2005 | 10.7 | 39.7 | 26.95 | 27.3 | 39.19 | 3.3 | 324.24 | 6.9 | 155.07 | 2.3 | 465.22 |
| 2006 | 11.4 | 42.1 | 27.08 | 28.4 | 40.14 | 3.5 | 325.71 | 6.8 | 167.65 | 3.3 | 345.45 |
| 2007 | 11.3 | 47.2 | 23.94 | 31.3 | 36.10 | 4.6 | 245.65 | 6.6 | 171.21 | 4.7 | 240.43 |
| 2008 | 11.6 | 51.2 | 22.66 | 34.1 | 34.02 | 5.6 | 207.14 | 6.9 | 168.12 | 4.6 | 252.17 |
| 2009 | 10.4 | 39.0 | 26.67 | 27.0 | 38.52 | 4.3 | 241.86 | 5.4 | 192.59 | 2.3 | 452.17 |
| 2010 | 9.6 | 39.5 | 24.30 | 28.5 | 33.68 | 4 | 240.00 | 4.7 | 204.26 | 2.4 | 400 |
| 2011 | 10.5 | 39.3 | 26.72 | 28.6 | 36.71 | 3.3 | 318.18 | 4.4 | 238.64 | 2.9 | 362.07 |
| 2012 | 10.4 | 39.0 | 26.67 | 27.5 | 37.82 | 3.8 | 273.68 | 5.1 | 203.92 | 2.6 | 400 |

$C_{11}$. Tourism – Travel Exchange – TE (billion €)  
$C$. Total of Invisible Revenues – IR (billion €)  
$C_{11}/C$ . Contribution of the TE in the balance of IR (%)  
$C_1$. Balance of Services – BS (billion €)  
$C_{11}/C_1$. Contribution of the TE in the BS (%)  
$C_2$. Balance of Income – BI (billion €)  
$C_{11}/C_2$. Relevance of the TE with the BI (%)  
$C_3$. Balance of Current Transfers - BCT  
$C_{11}/C_3$. Relevance of the TE with the BCT (%)  
$C_4$. Balance of Capital Transfers - BCAT  
$C_{11}/C_4$. Relevance of the TE with the BCAT (%)  
(source: BOG, 2014; own elaboration)

Additionally, for the visualization of these results the diagram 8 is constructed, presenting the annual change of tourist exchange in comparison to the invisible revenue's components. According to table 5 and to the diagram 8, the tourist exchange ($C_{11}$) at the period 2000-2012 presented the following performance: it contributed in an amount of 34-48% in the balance of services, it participated in a percentage of 22.74%-33.66% in the balance of invisible revenues and it is presented significantly greater than each of the other invisible revenues' components ($C_2$, $C_3$, $C_4$).

Also it seems that it shows a diachronically invariant performance, despite the effects of economic crisis started in the year 2009, having an average value of 10.51 billion € and approximately 6% annual fluctuations. This observation elects the importance of the foreign tourism as a developmental factor in the national economy of Greece, providing indications to the policy makers the country to focus on this aspect for implementing a stable national economic growth of low risk.

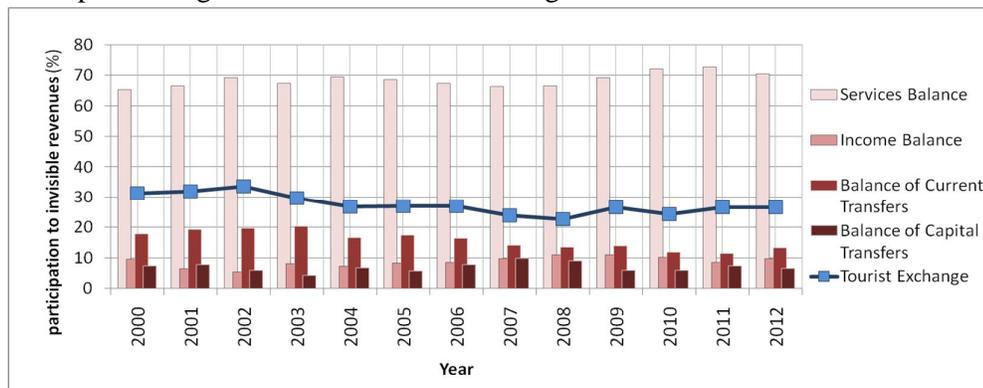

**Diagram 8:** Annual change of tourist exchange in comparison to the invisible revenue's components (source: table 5, own elaboration).

4.5.4. The contribution of Tourist Exchange in the Trade Balance

The foregoing analysis elected the obvious contribution of the tourist exchange in the national economy of Greece. Within this framework, this paragraph examines the contribution of the travel exchange in the coverage of the deficits of the payments' balance (BP). According to the Bank of Greece (BOG, 2014), the trade balance, which is coded in the account of payments with the letter *B*, is defined as the difference between the exports' revenue ($B_1$) minus imports expenditure ($B_2$) of the country, as it is shown in the following relation:

*Trade Balance (B) = Exports' Revenue ($B_1$) – Imports' Expenditure ($B_2$)*      (3)

Table 6 presents the diachronic change of the tourist exchange in comparison to the trade balance and its components, for the period 2000-2012.

**Table 6**

Diachronic change of the tourist exchange (TE) in comparison to the trade balance (TB) and its components, for the period 2000-2012

| Year | B | $B_1$ | $B_2$ | $C_{11}$ | $C_{11}/B$ | $C_{11}/B_1$ | $C_{11}/B_2$ |
|------|------|------|-------|-------|--------|--------|--------|
| 2000 | -21.9 | 11.1 | -33.0 | 10.1 | -46.12 | 90.99 | -30.61 |
| 2001 | -21.6 | 11.5 | -33.2 | 10.6 | -49.07 | 92.17 | -31.93 |
| 2002 | -22.7 | 10.4 | -33.1 | 10.3 | -45.37 | 99.04 | -31.12 |
| 2003 | -22.6 | 11.1 | -33.8 | 9.5 | -42.04 | 85.59 | -28.11 |
| 2004 | -25.4 | 12.7 | -38.1 | 10.3 | -40.55 | 81.10 | -27.03 |
| 2005 | -27.6 | 14.2 | -41.8 | 10.7 | -38.77 | 75.35 | -25.60 |
| 2006 | -35.3 | 16.2 | -51.4 | 11.4 | -32.29 | 70.37 | -22.18 |
| 2007 | -41.5 | 17.4 | -58.9 | 11.3 | -27.23 | 64.94 | -19.19 |
| 2008 | -44.0 | 19.8 | -63.9 | 11.6 | -26.36 | 58.59 | -18.15 |
| 2009 | -30.8 | 15.3 | -46.1 | 10.4 | -33.77 | 67.97 | -22.56 |
| 2010 | -28.3 | 17.1 | -45.4 | 9.6 | -33.92 | 56.14 | -21.15 |
| 2011 | -27.2 | 20.2 | -47.5 | 10.5 | -38.60 | 51.98 | -22.11 |
| 2012 | -19.6 | 22.0 | -41.6 | 10.4 | -53.06 | 47.27 | -25.00 |

$B$. Trade Balance – TB (billion €)  
$B_1$. Exports' Revenue – ER (billion €)  
$B_2$. Imports' Revenue – IMR (billion €)  
$C_{11}$. Tourist – Travel Exchange – TE (billion €)  

$C_{11}/B$. Relevance of the TE with the TB (%)  
$C_{11}/B_1$. Relevance of the TE with the ER (%)  
$C_{11}/B_2$. Relevance of the TE with the IMR (%)  
(source: BOG, 2014; own elaboration)

For the better comprehension of the relations presented in table 6 the diagram 9 is constructed, showing the diachronic evolution of the TE ($C_{11}$) in a line plot and of the TB ($B$) and its components ($B_1$, $B_2$) in bar plots. According to this diagram, the size of the tourist exchange presents an almost invariant performance diachronically (which is lying around the amount of 10.000 million €), which is comparable with the amount of the national exports that, however, present a progressive increase (from the amount of 11.000 million €, in 2000, to the amount of 22.000 million €, approximately, in 2012) in the last years.

Further, table 6 illustrates that the percentage relation of the tourist exchange ($C_{11}$) in comparison to the trade balance ($B$) is considerably high, lying between 26.42%-53.23%. This observation underlines the importance of the travel exchange to the coverage of the deficits of the trade balance of Greece and consequently to the final configuration of the balance of current transactions.

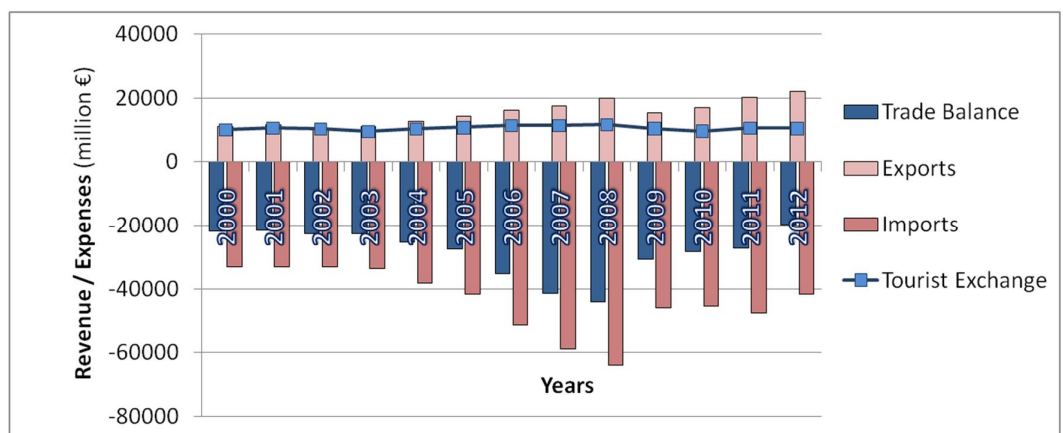

**Diagram 9:** Annual change of the Tourist Exchange in comparison to the Trade Balance and its components, in Greece (source: table 6, own elaboration).

Finally, diagram 9 illustrates the stability in the incoming flows of the tourist exchange in the country and also the resilience that this measure demonstrates at the current period of the economic crisis. Further, the information provided in table 6 and in diagram 9 elect the impressive observation that the economic crisis that started exhausting the country from the year 2009 led to the increase of the exports of goods, instead of decreasing them. This observation requires further examination in order to provide clearer insights, but is outside the purpose of this paper suggesting a topic of further research.

### 5. Empirical analysis

This section studies the diachronic variable of *tourism* or *travel exchange* (TE) and examines its contribution to the *balance of payments* account (BP), using empirical and statistical techniques. The variables that participated in the part of the empirical analysis are shown in table 7, where the variable YEAR is manipulated as an ordinal variable, while all the others as scale variables.

At first, an independent samples *t*-test is applied for the comparison of the means $\mu_{1, year \geq 2008}^{(i)}, \mu_{2, year < 2008}^{(i)}$, $i=,...,10$, of the available variables in table 7 and its results are shown in table 8. The analysis uses as cutting point the year 2008, aiming in detecting whether the performance of each variable has stayed inalterable from the economic crisis that started affecting the country from this year and after. The null hypothesis states the inequality of means ($H_o$: $\mu_1 \neq \mu_2$), while the alternative the equality ($H_1$: $\mu_1 = \mu_2$). Cases rejecting the null hypothesis imply that the examined variables were resilient to the effects of the economic crisis and thus they suggest factors of low risk for the economic development of the country.

**Table 7**
Variables participating in the empirical analysis (values in billion €)

| YEAR | GNP | TE | TB | ER | IMR | IR | BS | BI | BCT | BCAT |
|---|---|---|---|---|---|---|---|---|---|---|
| 2000 | 136.3 | 10.06 | -21.9 | 11.1 | -33 | 32.14 | 20.98 | 3.04 | 5.74 | 2.39 |
| 2001 | 146.4 | 10.58 | -21.6 | 11.5 | -33.2 | 33.12 | 22.08 | 2.1 | 6.34 | 2.6 |
| 2002 | 156.6 | 10.29 | -22.7 | 10.4 | -33.1 | 30.56 | 21.13 | 1.63 | 5.98 | 1.82 |
| 2003 | 172.4 | 9.5 | -22.6 | 11.1 | -33.8 | 31.83 | 21.43 | 2.58 | 6.42 | 1.39 |
| 2004 | 185.3 | 10.35 | -25.4 | 12.7 | -38.1 | 38.53 | 26.74 | 2.81 | 6.36 | 2.62 |
| 2005 | 193 | 10.73 | -27.6 | 14.2 | -41.8 | 39.73 | 27.25 | 3.27 | 6.88 | 2.32 |
| 2006 | 208.6 | 11.36 | -35.3 | 16.2 | -51.4 | 42.05 | 28.36 | 3.53 | 6.85 | 3.31 |
| 2007 | 223.2 | 11.32 | -41.5 | 17.4 | -58.9 | 47.18 | 31.34 | 4.56 | 6.61 | 4.67 |
| 2008 | 233.2 | 11.64 | -44 | 19.8 | -63.9 | 51.16 | 34.07 | 5.57 | 6.88 | 4.64 |
| 2009 | 231.1 | 10.4 | -30.8 | 15.3 | -46.1 | 38.98 | 26.98 | 4.28 | 5.38 | 2.33 |
| 2010 | 222.2 | 9.61 | -28.3 | 17.1 | -45.4 | 39.5 | 28.48 | 4.01 | 4.65 | 2.36 |
| 2011 | 208.5 | 10.51 | -27.2 | 20.2 | -47.5 | 39.3 | 28.61 | 3.32 | 4.44 | 2.93 |
| 2012 | 193.3 | 10.44 | -19.6 | 22 | -41.6 | 39.05 | 27.53 | 3.83 | 5.13 | 2.56 |

GNP=Gross National Product  
TE=Tourist – Travel Exchange  
TB=Trade Balance  
ER=Exports' Revenue  
IMR=Imports' Revenue  
IR= Total of Invisible Revenues  
BS= Balance of Services  
BI= Balance of Income  
BCT= Balance of Current Transfers  
BCAT= Balance of Capital Transfers  
(source: BOG, 2014; own elaboration)

According to table 8, the cases rejecting the null hypothesis refer to the variables of *tourist exchange* (TE), the *trade balance* (TB), the *imports' revenue* (IMR), the *total of invisible revenues* (IR), the *balance of services*, and to the *balance of capital transfers* (BCAT). The cases TE and BS imply the rigid economic basis of the sector of services in Greece and of its orientation to tourism.

**Table 8**

Independent samples *t*-test for the comparison of means $\mu_{1, year \geq 2008}^{(i)}$, $\mu_{2, year<2008}^{(i)}$ $^{(a),(b)}$, $i=,\ldots,10$

| | Equal variances: | Levene's Test for Equality of Variances | | t-test for Equality of Means | | | | | 95% C.I of the Difference | |
|---|---|---|---|---|---|---|---|---|---|---|
| | | F | Sig. | t | df | Sig.$^{(c)}$ | Mean Difference | Std. Error Difference | Lower | Upper |
| (*i*=1) GNP$^{(d)}$ | assumed | 2.461 | .145 | 2.664 | 11 | .022 | 39.935 | 14.989 | 6.945 | 72.925 |
| | n/a$^{(e)}$ | | | 3.047 | 10.933 | .011 | 39.935 | 13.107 | 11.066 | 68.804 |
| (*i*=2) TE$^{(f)}$ | assumed | .013 | .911 | -.008 | 11 | **.994** | -0.003 | 0.378 | **-0.834** | **0.828** |
| | n/a | | | -.007 | 7.651 | .994 | -0.003 | 0.392 | -0.913 | 0.908 |
| (*i*=3) TB$^{(g)}$ | assumed | .014 | .908 | -.588 | 11 | **.568** | -2.655 | 4.514 | **-12.591** | **7.281** |
| | n/a | | | -.560 | 7.353 | .592 | -2.655 | 4.739 | -13.753 | 8.443 |
| (*i*=4) ER$^{(h)}$ | assumed | .000 | 1 | 3.884 | 11 | .003 | 5.805 | 1.495 | 2.515 | 9.095 |
| | n/a | | | 3.861 | 8.475 | .004 | 5.805 | 1.503 | 2.372 | 9.238 |
| (*i*=5) IMR$^{(i)}$ | assumed | .313 | .587 | -1.582 | 11 | **.142** | -8.488 | 5.365 | **-20.296** | **3.321** |
| | n/a | | | -1.632 | 9.495 | .135 | -8.488 | 5.201 | -20.160 | 3.185 |
| (*i*=6) IR$^{(j)}$ | assumed | .510 | .490 | 1.442 | 11 | **.177** | 4.706 | 3.262 | **-2.474** | **11.885** |
| | n/a | | | 1.480 | 9.346 | .172 | 4.706 | 3.180 | -2.447 | 11.858 |
| (*i*=7) BS$^{(k)}$ | assumed | 3.039 | .109 | 2.045 | 11 | **.066** | 4.220 | 2.064 | **-0.323** | **8.763** |
| | n/a | | | 2.220 | 10.681 | .049 | 4.220 | 1.901 | 0.022 | 8.419 |
| (*i*=8) BI$^{(l)}$ | assumed | .068 | .800 | 2.520 | 11 | .028 | 1.262 | 0.501 | 0.160 | 2.364 |
| | n/a | | | 2.562 | 9.100 | .030 | 1.262 | 0.493 | 0.150 | 2.374 |
| (*i*=9) BCT$^{(m)}$ | assumed | 2.530 | .140 | -2.929 | 11 | .014 | -1.102 | 0.376 | -1.929 | -0.274 |
| | n/a | | | -2.437 | 4.860 | .060 | -1.102 | 0.452 | -2.273 | 0.070 |
| (*i*=10) BCAT$^{(n)}$ | assumed | .000 | .990 | .576 | 11 | **.576** | 0.324 | 0.563 | **-0.914** | **1.562** |
| | n/a | | | .581 | 8.856 | .576 | 0.324 | 0.558 | -0.942 | 1.590 |

(a) Null hypothesis: $\mu_1 \neq \mu_2$     (e) n/a=not assumed     (i) Imports' Revenue     (m) Balance of Current
(b) Cutting point at the year 2008     (f) Tourist – Travel Exchange     (j) Total of Invisible Revenues     Transfers
(c) (2-tailed)     (g) Trade Balance     (k) Balance of Services     (n) Balance of Capital
(d) Gross National Product     (h) Exports' Revenue     (l) Balance of Income     Transfers
    - cases in bold reject the null hypothesis

Further, the other cases (TB, IMR, IR, BCAT) they are probably illustrating a framework of fixed external needs in Greece, which are being served by imports, in conjunction with the marine specialization of the country, which constitutes a solid factor of its national economy. However, the most significant case, among those that reject the null hypothesis, corresponds to this of tourist exchange, probably highlighting the best resilience of this variable to the economic crisis.

Next, a correlation analysis is applied on the data of the table 7, in order to detect structural similarities between variables. In this case, the Gross National product (GNP) was chosen as a reference variable, detecting which other variables scale diachronically in a similar to the GNP way. The results of the correlation analysis are shown in table 9, where it can be observed that the correlation *r*(GNP,TE)=.427 is insignificant, implying a possibility of 14.6% to be a result of chance. This, in conjunction with the picture shaped in diagram 7, where the diachronic curve of the GNP is significantly affected by the economic crisis, while this of the TE is not, it validates the resilience of the tourism exchange variable.

Another interesting observation drafted from the correlation analysis of table 9 is the negative significant and high correlation *r*(GNP,IMR)=-.858$^{**}$, which first implies these variables scale similarly but inversely in the time line. Further, this result implies that imports generally operate as a brake in the increase of the GNP of the country and that their contribution in the formulation of the diachronic pattern of the GNP is important.

**Table 9**
Pearson's bivariate correlations between GNP and component of the Balance of Payments

|     |          | TE    | TB       | ER       | IMR       | IR        | BS        | BI        | BCT   | BCAT     |
|-----|----------|-------|----------|----------|-----------|-----------|-----------|-----------|-------|----------|
| GNP | $r_{xy}$(a) | .427  | -.749**  | .727**   | -.858**   | .831**    | .891**    | .834**    | -.099 | .561*    |
|     | Sig.     | .146  | .003     | .005     | .000      | .000      | .000      | .000      | .749  | .046     |
|     | N        | 13    | 13       | 13       | 13        | 13        | 13        | 13        | 13    | 13       |
| TE  | $r_{xy}$ |       | -.757**  | .428     | -.749**   | .760**    | .670*     | .520      | .532  | .854**   |
|     | Sig.     |       | .003     | .144     | .003      | .003      | .012      | .068      | .061  | .000     |
|     | N        |       | 13       | 13       | 13        | 13        | 13        | 13        | 13    | 13       |
| TB  | $r_{xy}$ |       |          | -.428    | .937**    | -.891**   | -.819**   | -.781**   | -.398 | -.861**  |
|     | Sig.     |       |          | .144     | .000      | .000      | .001      | .002      | .178  | .000     |
|     | N        |       |          | 13       | 13        | 13        | 13        | 13        | 13    | 13       |
| ER  | $r_{xy}$ |       |          |          | -.718**   | .723**    | .809**    | .738**    | -.350 | .564*    |
|     | Sig.     |       |          |          | .006      | .005      | .001      | .004      | .242  | .044     |
|     | N        |       |          |          | 13        | 13        | 13        | 13        | 13    | 13       |
| IMR | $r_{xy}$ |       |          |          |           | -.967**   | -.945**   | -.888**   | -.171 | -.883**  |
|     | Sig.     |       |          |          |           | .000      | .000      | .000      | .576  | .000     |
|     | N        |       |          |          |           | 13        | 13        | 13        | 13    | 13       |
| IR  | $r_{xy}$ |       |          |          |           |           | .977**    | .895**    | .239  | .887**   |
|     | Sig.     |       |          |          |           |           | .000      | .000      | .432  | .000     |
|     | N        |       |          |          |           |           | 13        | 13        | 13    | 13       |
| BS  | $r_{xy}$ |       |          |          |           |           |           | .878**    | .066  | .812**   |
|     | Sig.     |       |          |          |           |           |           | .000      | .831  | .001     |
|     | N        |       |          |          |           |           |           | 13        | 13    | 13       |
| BI  | $r_{xy}$ |       |          |          |           |           |           |           | .013  | .733**   |
|     | Sig.     |       |          |          |           |           |           |           | .968  | .004     |
|     | N        |       |          |          |           |           |           |           | 13    | 13       |
| BCT | $r_{xy}$ |       |          |          |           |           |           |           |       | .315     |
|     | Sig.     |       |          |          |           |           |           |           |       | .295     |
|     | N        |       |          |          |           |           |           |           |       | 13       |

**. Correlation is significant at the 0.01 level (2-tailed).           (a) Pearson Correlation
*. Correlation is significant at the 0.05 level (2-tailed).            (b) Sig. (2-tailed)

Whether using the tourism exchange (TE) as a reference variable, we can see that it is correlated with the most cases of the other variables. The insignificant correlation $r$(TE,ER)=.428 interprets that value is 14.4% probable to be a result of chance, while for the insignificant coefficients $r$(TE,BI)=.520 and $r$(TE,BCT)=.532 the respective probabilities are 6.8% and 6.1%. On the other hand, the significant coefficients $r$(TE,TB)= -.757**, $r$(TE,IMR)=-.749**, $r$(TE,IR)=.760**, $r$(TE,BS)=.670* and $r$(TE,BCAT)=.854** illustrate the representative role of TE in the diachronic evolution of the balance of payments, a fact that comes to an agreement with the findings of the foregoing descriptive analysis.

The final step of the empirical analysis regards the application of fitting curves on the diachronic data of each variable shown in table 7, in order to detect statistically significant slopes and thus evolving trends per case. In particular, the curve fitting procedure quantifies a linear patter $y=bx+c$ that describes the diachronic fluctuation of the examined variables. Diagram 10 shows the diachronic change of the elements composing the Balance of Payment, which have significant slopes (linear coefficients beta – $b$), along with the value of the coefficient of determination $R^2$ (R-square), which describes the amount of variation of the response variable's ($y$) data that is captured by each linear fitting.

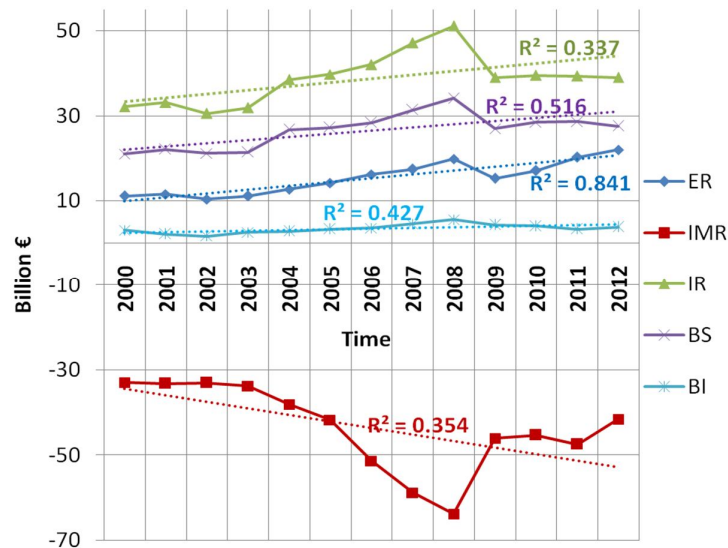

**Diagram 10:** Annual change of the Balance of Payment's components (that they are included in table 7) with significant linear coefficients beta (*b*).

The absence of the variable TE (tourism exchange) among the significant fittings shown in diagram 10 implies, on the one hand, the resilience of this variable to externalities, highlighting its stable performance on the Greek economy. This comes to an agreement with the findings of the foregoing analysis, which outlined that tourism constitutes diachronically one of the basic components of the national economy of Greece.

On the other hand, the failure of the TE variable to shape a significant growth pattern that is represented by a statistically significant positive slope beta (*b*) elects a topic of concern for the stakeholders of the Greek tourism policy. Taking into consideration the variety of the physical resources in Greece, which favor the development of many forms of tourism and in many geographical places, this result is probably illustrating a deficiency in the Greek policy to employ an effective plan for the development of Greek tourism, with the simultaneous exploitation of the endogenous tourism dynamics.

In other words, the picture between the significant proportion of the touristic revenues in Greece and their incapability to present a growing pattern is quite contradictive, implying that the tourism development in Greece is probably constrained by its own complexity concluding to evolve singularly through a fuzzy endogenous mechanism and not through an organized sustainable plan.

1. Conclusions

The foregoing analysis outlined that tourism constitutes diachronically one of the basic components of the national economy of Greece. Its effects in the configuration of economic framework of the country are determinative, being sprawled in many levels and sectors of the national and regional economic activity. The previous consideration was verified by the empirical analysis, evincing that the contribution of tourism in the economic growth of Greece is determinative, both through the macroeconomic and the micro-economic perspectives.

This paper also outlined that one of the major measurable economic parameters of tourism is the travel or tourist exchange, which constitutes the most important account, in terms of size, among the components of the invisible revenues, which it contributes in the inflow of substantial (non accountant) money in the domestic market and sequentially in the reduction of the deficit of the balance of payments. Also, the foregoing empirical consideration emerged the diachronic invariance of the tourism income, which demonstrated its resilience at the period of economic crisis, implying that the tourism constitutes a promising investment of "low risk" for the economic growth of country and of its exit of the crisis.

It is worth to be noted that tourism may suggest for the geographically isolated regions of country, such as are the islands and the mountainous regions, an activity that is capable to preserve their demographic ageing, contributing in the staying of the young and thus economically active people in such regions. With this way, the Greek tourism can operate as a basic developmental lever, not only in national level, but also in regional level, contributing in the economic strength of the weak regions and in the equalization of the regional problem.

The foregoing analysis did not illustrate that the Greek tourism suggests a healthy productivity mechanism that it lacks of structural problems; it only elected its dynamics to constitute a stable developmental axis and a channel for the possible exit of country from the economic crisis. For the increase of the effectiveness of tourism, whether considering it as an economic activity, it is necessary for the policy makers of the country to take into account and to focus on the exploitation of all the factors that were described in Section 2. Thus, it is necessary to focus on tourists that spend greater daily amounts and to attract them to stay in their tourism destination for longer time periods, in order to increase the total tourist expenditure per visitor.

Next, managing the tourism seasonality suggests another issue that should be addressed, both through the elongation of the tourist period for the so-called "mass" or "holidays" tourism, and through the growth of alternative forms of tourism. This is expected to contribute in a better geographic distribution of tourism in Greece, which today is characterized by an excessive agglomeration in certain regions, such as Corfu, Northern Crete, Rhodes and Chalkidiki, inducing obvious consequences to the tourism carrying capacity of these places.

Finally, the foregoing analysis outlined a contradictive picture between the share of tourism in the Greek economy and its evolving pattern, highlighting that the tourism development in Greece is probably constrained by its own complexity, which forces tourism to evolve singularly through a fuzzy endogenous mechanism and not through an organized sustainable plan. This instructs that tourism policy should be focused on the increase of the tourism multipliers through the exploitation and incorporation in the tourist consumption of the locally produced products, a fact that it is expected to reinforce the endogenous developmental dynamics of the regions and to upgrade the quality of the tourism goods.